\pdfoutput=1

\documentclass[11pt,twoside,a4paper,cmspaper,final,collab]{cms-tdr}
\def\svnVersion{cc18f21-D}\def\svnDate{2019/07/11}\def\cmsCernNoTag{CERN-EP-2019-040}\def\cmsCernDate{\today}\def\cmsMessage{"Published in the Journal of High Energy Physics as \href{http://dx.doi.org/10.1007/JHEP06(2019)143}{\doi{10.1007/JHEP06(2019)143}.}"}

\begin{document}\cmsNoteHeader{SUS-17-011}

\hyphenation{had-ron-i-za-tion}
\hyphenation{cal-or-i-me-ter}
\hyphenation{de-vices}
\RCS$HeadURL: svn+ssh://svn.cern.ch/reps/tdr2/papers/SUS-17-011/trunk/SUS-17-011.tex $
\RCS$Id: SUS-17-011.tex 494396 2019-05-08 03:22:58Z areinsvo $

\newcommand{\ff}{\ensuremath{f \! f}\xspace}

\cmsNoteHeader{SUS-17-011}

\title{Search for supersymmetry in final states with photons and missing transverse momentum in proton-proton collisions at 13\TeV}

\date{\today}

\abstract{
        Results are reported of a search for supersymmetry in final states
        with photons and missing transverse momentum in proton-proton
        collisions at the LHC. The data sample
        corresponds to an integrated luminosity of 35.9\fbinv collected at a
        center-of-mass energy of 13\TeV using the CMS detector.
        The results are interpreted in the context of models of
        gauge-mediated supersymmetry breaking. Production cross section
        limits are set on gluino and squark pair
        production in this framework. Gluino masses below 1.86\TeV and
        squark masses below 1.59\TeV are excluded at 95\% confidence level.
        }

\hypersetup{
pdfauthor={CMS Collaboration},
pdftitle={Search for supersymmetry in final states with photons and missing transverse momentum in proton-proton collisions at 13 TeV},
pdfsubject={CMS},
pdfkeywords={CMS, physics, supersymmetry, photons and missing transverse momentum}}

\maketitle

\section{Introduction}
\label{sec:Introduction}
Supersymmetry (SUSY) and the
Minimal Supersymmetric Standard 
Model~\cite{Nappi:1982bc, Dine:1982cd,Alvarez:1982de, Dine:1993yw, Dine:1994vc,Dine:1995ag}
are extensions of the standard model (SM) that
provide explanations for several outstanding 
issues with the SM. In particular, SUSY addresses
the large quantum corrections to the mass term
in the Higgs potential~\cite{Barb} and provides a viable
dark matter candidate~\cite{DM1,DM2}. Models with
general gauge-mediated (GGM) SUSY 
breaking~\cite{Dimopoulos:1996fg,Martin:1997hi, Poppitz:1997gh, Meade:2008wd,Buican:2008ws, Abel:2009ve, Carpenter:2009jk, Dumitrescu:2010ef}
have the additional benefit of naturally suppressing
flavor violations in the SUSY sector. GGM models
can have a wide range of features but typically result in final states
that include the gravitino (\PXXSG) as the lightest supersymmetric particle (LSP).
The next-to-lightest supersymmetric particle (NLSP) in these models
is often taken to be a neutralino (\PSGczDo), which is a mixture 
of the bino, neutral wino, and neutral higgsinos.  
The conservation of
$R$ parity~\cite{FF} implies that the gravitino is stable and remains undetected.
Therefore, proton-proton (\Pp\Pp) collisions
that produce SUSY particles will have an imbalance in the
total observed transverse momentum, referred to as missing
transverse momentum \ptvecmiss and defined
as the negative vector sum
of the transverse momenta of all visible particles in an event.
Its magnitude is referred to as \ptmiss.
If the composition of the neutralino NLSP is primarily bino-like, its main decay will be to a
gravitino and a photon (\Pgg), resulting in final
states with significant missing transverse momentum and one or more photons.

This paper presents a search for GGM SUSY in final states involving two
photons and missing transverse momentum. The data sample, corresponding to an
integrated luminosity of 35.9\fbinv of $\Pp\Pp$ collisions at a
center-of-mass energy $\sqrt{s} = 13\TeV$,
was collected with the CMS detector in 2016.
The analysis described here achieves a substantial improvement in sensitivity
compared to the search performed by the CMS Collaboration on the
smaller 2015 data set~\cite{CMS:2016_anal} and is comparable in sensitivity
to similar searches from the ATLAS Collaboration~\cite{ATLAS:2016aa,ATLAS:2018_atlas}.

\begin{figure}[htbp]
    \centering
    \includegraphics[width=0.45\textwidth]{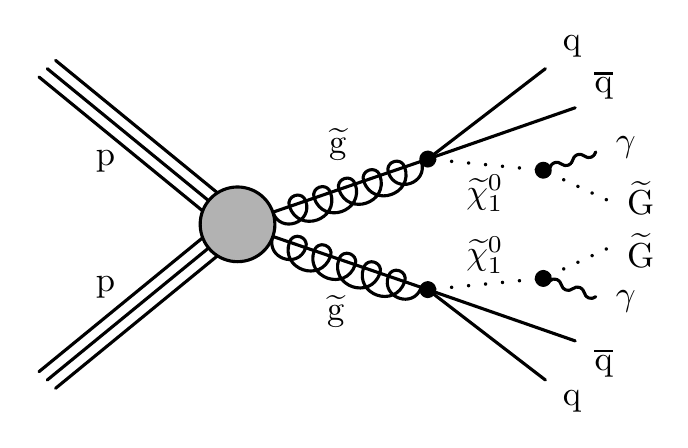}
    \includegraphics[width=0.45\textwidth]{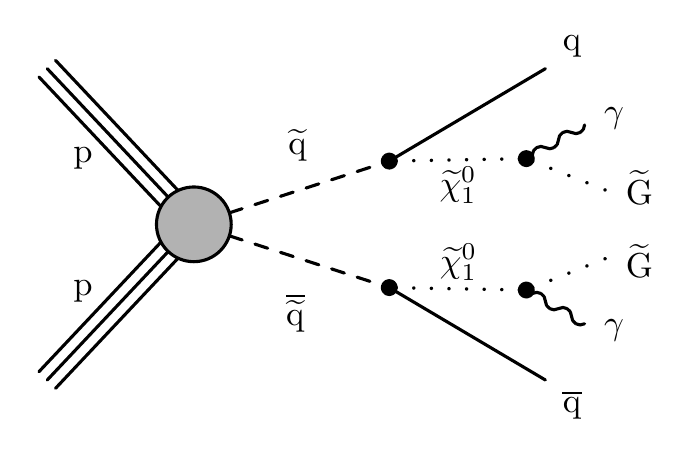}
    \caption{Diagrams showing the production of signal events in the collision
        of two protons (\Pp). In gluino
        (\PSg) pair production in the T5gg simplified model (left), the gluino
        decays to a quark-antiquark pair (\cPq\cPaq) and a neutralino (\PSGczDo). In
        squark (\PSq) pair production in the T6gg simplified model (right), the
        squark decays to a quark and a neutralino. In both cases, the
        neutralino subsequently decays to a photon (\Pgg) and a gravitino (\PXXSG).}
    \label{fig:gluinoSquarkDecay}
\end{figure}

Two simplified model frameworks~\cite{bib-sms-1,bib-sms-2,bib-sms-3,bib-sms-4,expPaper} 
are used for the interpretation of the results.
The T5gg model assumes gluino (\PSg) pair production and the T6gg model assumes
squark (\PSq) pair production. The models assume a 100\% branching fraction
for the gluinos and squarks to decay as shown in Fig.~\ref{fig:gluinoSquarkDecay}.
The squarks in the T6gg model can be either first or second generation.
We assume a 100\% branching fraction for the NLSP neutralino to decay
to a nearly massless gravitino and a photon, $\PSGczDo \to \PXXSG\Pgg$,
resulting in characteristic events with large \ptmiss and two photons.
In order for the analysis to be as model independent as possible,
we choose not to define the signal region using any hadronic variables such as jet multiplicity
or the scalar sum of the transverse momentum of the jets.

Standard model processes such as direct diphoton production or events with
jets produced through the strong interaction, referred to as
quantum chromodynamics (QCD) multijet events, can result in events with two photons.
If the hadronic activity in the event is
poorly measured, these processes
can mimic the signal topology even though they lack genuine \ptmiss.
For the case of QCD multijet events, there may be real photons in the event,
or jets rich in electromagnetic (EM) energy that are misreconstructed as photons.
Events with genuine \ptmiss also contribute to the composition of the candidate sample.
These events are mainly from $\PW\Pgg$ and
\PW+jet(s) production, where an electron is misidentified as a photon in $\PW\to \Pe\PGn$ decays.
A smaller background arises from $\cPZ\Pgg\Pgg$ events where the {\cPZ} boson decays to two
neutrinos, $\cPZ\to\cPgn\cPagn$.

\section{Detector, data, and simulated samples}
\label{sec:DataSimSamples}

The central feature of
the CMS apparatus is a superconducting solenoid of 6\unit{m} internal diameter,
providing a magnetic field of 3.8\unit{T}. Within the solenoid volume are a
silicon pixel and strip tracker covering the pseudorapidity region $\abs{\eta} <
2.5$, as well as a lead tungstate crystal electromagnetic calorimeter (ECAL),
and a brass and scintillator hadron calorimeter (HCAL), each composed of a
barrel and two endcap regions and covering the range $\abs{\eta} < 3.0$. Forward
calorimeters extend the coverage up to $\abs{\eta} < 5.0$. Muons are measured in
gas-ionization detectors embedded in the steel flux-return yoke outside the
solenoid and cover the range $\abs{\eta} < 2.4$. A more detailed description of the
CMS detector, together with a definition of the coordinate system used and the
relevant kinematic variables, can be found in Ref.~\cite{Chatrchyan:2008zzk}.

Events of interest are selected using a two-tiered trigger system~\cite{Khachatryan:2016bia}.
The first level is composed of custom hardware processors and uses
information from the calorimeters and muon detectors to select events at a rate of
around 100\unit{kHz}. The second level, known as the high-level trigger,
consists of a farm of processors running a version of the full event reconstruction
software optimized for fast processing. This trigger reduces the event rate to around 1\unit{kHz}
before data storage. This analysis used a diphoton trigger to collect the data.
The trigger requires a leading (subleading) photon with transverse momentum $\pt > 30$ (18)\gev,
and a combined invariant mass $m_{\Pgg\Pgg} > 95$\gev. The photons are
also required to pass isolation and cluster shape requirements.

Monte Carlo (MC) simulations are used for several purposes in this analysis. Simulations
of the signal processes are used to determine signal efficiencies;
background process simulation is used for validation of the analysis performance
and to model the contribution from $\cPZ\Pgg\Pgg\to\cPgn\cPagn\Pgg\Pgg$
events. The event generator \MGvATNLO~2.3.3~\cite{Alwall:2014hca}
is used to simulate the signal samples at leading order.
The background samples are generated at next-to-leading order using \MGvATNLO~2.4.2.
For both signal and background processes, the parton showering, hadronization,
SUSY particle decays, multiple-parton interactions, and the underlying event are described by the
\PYTHIA 8.212~\cite{Sjostrand:2007gs} program with the CUETP8M1~\cite{CMS:CUETP8M1_cite} generator tune.
The signal samples are generated with either two gluinos or two squarks and up to two additional
partons in the matrix element calculation.
The parton distribution
functions (PDFs) are obtained from the NNPDF3.0~\cite{Ball:2014uwa} set. For the background
processes, the detector response is simulated using
\GEANTfour~\cite{Agostinelli:2002hh}, while the CMS fast
simulation~\cite{Abdullin:2011zz, Giammanco:2014bza} is used for the signal events.
For both signal and background simulated events,
additional $\Pp\Pp$ interactions (pileup) are generated with \PYTHIA
and superimposed on the primary collision process.
The simulated events are reweighted to match the pileup distribution observed in data.

The signal events were generated using the T5gg and T6gg simplified models and
are characterized by the masses of the particles in the decay chain. For the
gluino (squark) mass we simulate a range of values from 1.4 to 2.5 (1.2 to
2.0)\TeV in steps of 50\GeV. These mass ranges were selected to overlap and
expand upon the mass ranges excluded by previous
searches~\cite{ATLAS:2016aa,CMS:2016_anal}.
The neutralino masses range from 10\GeV up to the mass of the gluino or squark.
The cross sections are calculated at next-to-leading-order
(NLO) accuracy including the resummation of soft gluon emission
at next-to-leading-logarithmic (NLL) accuracy~\cite{Beenakker:1996ch,Kulesza:2008jb,Kulesza:2009kq, Beenakker:2009ha,Beenakker:2011fu},
with all the unconsidered sparticles assumed to be heavy and decoupled.
The uncertainties in the cross sections are calculated as
described in Ref.~\cite{Borschensky:2014cia}.

\section{Event selection}
\label{sec:EventSelect}
Photon, electron, muon, charged and neutral hadron candidates are
reconstructed with the particle-flow event algorithm~\cite{CMS-PRF-14-001},
which reconstructs particles based on information from all detector subsystems.
The energy of photons is directly obtained from the ECAL measurement.
The energy of electrons is determined from a combination of the electron momentum
at the primary interaction vertex as determined by the tracker,
the energy of the corresponding ECAL cluster, and the energy sum of all
bremsstrahlung photons spatially compatible with originating from the electron track.
The energy of muons is obtained from the curvature of the corresponding track.
The energy of charged hadrons is determined from a combination of their momentum
measured in the tracker and the matching ECAL and HCAL energy deposits,
corrected for zero-suppression effects and for the response function of the
calorimeters to hadronic showers. Finally, the energy of neutral hadrons
is obtained from the corresponding corrected ECAL and HCAL energy.

Photon candidates are required to satisfy a series of identification criteria
to ensure a high purity~\cite{Khachatryan:2015iwa}. The shape of the energy deposit in the ECAL
must be consistent with that of an EM shower, and
the amount of energy present in the corresponding region of the HCAL must not exceed 5\% of the
ECAL energy, since EM showers are expected to be contained almost
entirely within the ECAL. To ensure high trigger efficiency,
we require all photons to satisfy $\pt > 40$\GeV.
Because the SUSY signal models used in this analysis produce photons primarily
in the central region of the detector and because the magnitude of the
background increases considerably at high $\abs{\eta}$, we consider only
photons within the barrel fiducial region of the detector ($\abs{\eta} < 1.44$).

To suppress quark and gluon jets that mimic photons, photon candidates
are required to be isolated from other reconstructed particles. Separate
requirements are made on the scalar \pt sums of charged and neutral
hadrons and EM objects in a cone of radius
$\Delta R \equiv \sqrt{\smash[b]{ (\Delta\eta)^{2} + (\Delta\phi)^{2}}} \equiv 0.3$ around the photon candidate. Each \pt
sum is corrected for the effect of pileup, and in each case the momentum of the photon candidate itself is
excluded. We further require that the photon candidate has no pixel detector track
seed, to distinguish the candidate from an electron.

For the purpose of defining the various control regions used in the analysis,
we apply an additional set of selection criteria. A misidentified ``fake'' photon ($f$) is
defined as a photon candidate that satisfies looser requirements on
photon isolation and neutral-hadron isolation and  fails
either the shape requirement for the ECAL clusters or the charged-hadron
isolation requirement. In order to ensure that misidentified
photons do not differ too much from our photon selection, upper limits are
applied to both the charged-hadron isolation and cluster shape requirements.
Importantly, because of the large amount of hadronic activity expected in our SUSY signal events,
it is possible that real photons from the decay of a neutralino could fail the
charged-hadron isolation requirement and therefore fall into the misidentified photon category.
In order to avoid this potential signal contamination from SUSY events in the control regions,
we additionally require
that misidentified photons satisfy $\RNINE < 0.9$, where $\RNINE$ is defined as the ratio
of the energy deposited in a 3{$\times$}3 array of ECAL crystals to
the total energy in the cluster~\cite{Khachatryan:2015iwa}. Real photons have values
of $\RNINE$ close to unity, so by requiring $\RNINE < 0.9$ we ensure that real photons from
a possible SUSY signal will not enter our control regions.

Because of the similarity of the ECAL response to electrons and photons, $\cPZ\to\Pe\Pe$
events are used to measure the
photon identification efficiency. The selection of electron candidates is
identical to that of photons, with the exception that the candidate is required
to be matched to a pixel detector seed consistent with a track, to ensure that the electron
selection is orthogonal to that of photons.
The photon efficiency is measured via the tag-and-probe method~\cite{Khachatryan:2015iwa}.
The ratio of the observed to simulated efficiency is found to be consistent
with unity and independent of \pt and $\eta$.
The efficiency of the pixel detector seed veto for photons
is measured in $\cPZ\to\PGm\PGm\Pgg$ events and is found to
agree between data and simulation.

Events are then assigned to one of four mutually exclusive categories depending
on the selection of their highest \pt EM objects: $\Pgg\Pgg$,
$\Pe\Pe$, \ff, and $\Pe\Pgg$. The two
EM objects are required to be separated by $\Delta R > 0.6$. Finally, because of the
trigger requirements described in Section~\ref{sec:DataSimSamples},
the invariant mass of the two
EM objects is required to be greater than 105\gev.

In addition to the requirements already described, any event with a muon
satisfying $\pt > 25\gev$ and $\abs{\eta} < 2.4$ as well as track quality
and isolation requirements is vetoed.
Similarly, we veto events with any additional electrons satisfying
$\pt > 25\gev$, $\abs{\eta} < 2.5$, and signal shape and isolation
requirements.

Events in the candidate $\Pgg\Pgg$ sample are divided into the low-\ptmiss
control region ($\ptmiss < 100$\GeV) and the high-\ptmiss signal region
($\ptmiss > 100$\GeV). The signal region is further divided into
six \ptmiss bins that were chosen such that there is a
sufficient number of events from the \ff control sample in each bin.

\section{Estimation of backgrounds}
\label{sec:EstBkg}

As described in Section~\ref{sec:Introduction}, there are three primary backgrounds 
to this analysis. QCD processes such as multijet production 
can emulate the signal topology if the hadronic activity 
in the event is mismeasured. A second background arises 
from electroweak (EWK) processes that have genuine \ptmiss from the production of neutrinos.
There is also a small contribution from $\PZ\Pgg\Pgg\to\Pgg\Pgg\cPgn\cPagn$ events.

The contribution from the QCD background is estimated from the observed data using
the \ff control sample. The ratio of the event yield in the candidate
$\Pgg\Pgg$ sample to that in the \ff sample
is constructed as a function of \ptmiss. More \ff events are observed
at high \ptmiss relative to the $\Pgg\Pgg$ sample. Different functional
forms were investigated to model the \ptmiss dependence, and an exponential
function was found to describe the data the best.
We fit the $\Pgg\Pgg$ to \ff ratio in the
$\ptmiss < 100$\GeV control region. The predicted number of
QCD background events ($N_{\mathrm{QCD}}^i$) in bin $i$ of
the signal region is then given by the following equation, where
$N_{\ff}^{i}$ is the number of observed \ff events and
$g_{\mathrm{ave}}^{i}$ is the average value of the fit function $g(\ptmiss)$
in that bin:
\begin{equation}
N_{\mathrm{QCD}}^{i} = g_{\mathrm{ave}}^{i} \ N_{\ff}^{i}
\end{equation}

In order to set a systematic uncertainty on the method,
we derive a second QCD background prediction by noting that
the \ptmiss distribution of the \ff control sample is dependent on
the $\RNINE$ requirement on the misidentified photons.
An alternate \ff control sample is built using photon candidates that satisfy
all of the requirements for misidentified photons as
outlined in Section~\ref{sec:EventSelect}, with
the exception that the $\RNINE$ requirement is reversed.
In the $\ptmiss < 100$\GeV control region, we perform an exponential fit
to the ratio of the event yield in the high-$\RNINE$ \ff sample to that of the
nominal, low-$\RNINE$ \ff sample.
This function ($h(\ptmiss)$) represents
the correction required to account for the effect of the $\RNINE$ selection
on the \ptmiss distribution. The size of the correction is between
20 and 40\% in the $\ptmiss > 100$\GeV signal region.
Multiplying the number of low-$\RNINE$ \ff events
observed in the signal region by this function gives a proxy high-$\RNINE$ \ff sample.
\begin{equation}
N_{\mathrm{proxy}}^{i} = h_{\mathrm{ave}}^{i} \ N_{\ff}^{i}
\end{equation}

For $\ptmiss < 100$\GeV, the ratio of the \ptmiss distribution in the $\Pgg\Pgg$ sample to that of the
proxy \ff sample is fit to a constant $C$. 
We multiply this constant value by the proxy \ff yield in the signal region
to get a second prediction for the QCD background in bin $i$.
\begin{equation}
N_{\mathrm{QCD}}^{i} = C \ N_{\mathrm{proxy}}^{i}
\end{equation}

The two background estimation methods give values that are consistent
within the uncertainties. All three of the fits used in the two methods
are found to represent the data well in the low-\ptmiss control region.
Several studies were performed to verify the procedure, including using a
mixed-$\RNINE$ \ff sample with one misidentified photon
satisfying $\RNINE > 0.9$ and one satisfying $\RNINE < 0.9$
to confirm that the exponential fit continues to accurately describe the mixed-$\RNINE$ \ff to nominal \ff
ratio in the high-\ptmiss signal region. As an additional check, a control sample
with one photon and one misidentified photon
was used as a proxy for the $\Pgg\Pgg$ candidate sample in a closure test of the method 
up to $\ptmiss = 250$\GeV. At larger values of \ptmiss, 
there is the potential for signal contamination in the $\Pgg f$ control sample.

Another background for this analysis comes from EWK processes with genuine
\ptmiss. This background primarily involves $\PW\Pgg$ and {\PW}+jets events where the $\PW$ decays to an
electron and a neutrino and the electron is misidentified as photon. This leads to
final states with photons and significant \ptmiss.
To obtain an estimate of the EWK background in the signal region,
the mass peaks from the \PZ boson in the $\Pe\Pe$ control sample and 
the $\Pe\Pgg$ control sample are modeled using an extended likelihood fit for
the signal plus background hypothesis. The rate at which electrons are misidentified as photons 
($f_{\Pe \to \Pgg}$) is calculated using the signal fit integrals $N_{\Pe\Pgg}$ and $N_{\Pe\Pe}$
for each sample. These can be expressed in terms of the number of true
\PZ bosons, $N_{\PZ}^{\mathrm{True}}$: $N_{\Pe\Pe} = (1-f_{\Pe \to \Pgg})^2 N_{\PZ}^{\mathrm{True}}$ 
and $N_{\Pe\Pgg} = 2 f_{\Pe \to \Pgg} (1-f_{\Pe \to \Pgg})N_{\PZ}^{\mathrm{True}}$.
The factor of 2 in $N_{\Pe\Pgg}$ occurs because either electron in the event could be 
misidentified as a photon. 

Taking the ratio of 
these two values, we find that the misidentification rate is given by 
$f_{\Pe \to \Pgg} = N_{\Pe\Pgg}/(2N_{\Pe\Pe} + N_{\Pe\Pgg})$.
The misidentification rate is calculated as a function of several kinematic variables,
including the vertex multiplicity and \ptmiss of the event and the \pt of the EM objects.
A 30\% uncertainty is applied to cover the observed dependencies.
The final EWK background prediction is given by scaling the number of
events in the $\Pe\Pgg$ control sample by the factor
$f_{\Pe\Pgg \to \Pgg\Pgg} = f_{\Pe \to \Pgg}/ ( 1 - f_{\Pe \to \Pgg} )= (2.6 \pm 0.8)\%$.

The irreducible $\PZ\Pgg\Pgg$ background is modeled via simulation.
A 50\% uncertainty is applied to conservatively cover the effects from
the statistical uncertainty of the MC sample, the PDF uncertainty in the cross section, 
NNLO corrections in the simulation, and any other sources of potential mismodeling. 

\section{Sources of systematic uncertainty}
\label{sec:SysUncert}

Systematic uncertainties are calculated for each contribution
to the total background prediction. In addition,
systematic uncertainties are assigned for
the signal efficiency and the integrated luminosity. The
value of each uncertainty and the method used to calculate
it are described below.

The largest uncertainties
in the background prediction come from uncertainties associated with
the QCD background estimate.
The magnitude of each uncertainty is shown in
Table~\ref{tab:QCDerrors} for the six signal bins.
The statistical uncertainty
from the \ff control sample
ranges from 7 to 79\% in the signal region.
The uncertainty obtained from propagating the errors in the
fit parameters to the final prediction is between 2 and 5\%.
Finally, as described in Section~\ref{sec:EstBkg}, a systematic
uncertainty in the fitting procedure is calculated by
comparing the primary prediction to the cross check prediction
derived using the high-$\RNINE$ \ff sample. The systematic
uncertainty is taken as the difference between the two methods
or the uncertainty in that difference, whichever is larger,
and ranges between 10 and 83\% in the signal region.

\begin{table}[ht]
    \centering
     \topcaption{Event yield and statistical and systematic uncertainties (in numbers of events)
of the QCD background estimation for each signal \ptmiss bin for 35.9\fbinv of data at 13\TeV.}
     \centering
     \begin{tabular}{ c  c  c  c  c  c c}
     \hline
     \ptmiss bin (\GeVns{}) & Expected QCD &  Stat. uncert.   & Fit uncert. & Cross check uncert. \\
     \hline
     $100-115$  & 99.0 & $+7.2, -6.7$ & $\pm 1.8 $& $\pm 9.9 $\\
     $115-130$  & 32.8 & $+4.2, -3.7$ & $\pm 0.7 $& $\pm 5.5 $\\
     $130-150$  & 18.8 & $+3.2, -2.7$ & $\pm 0.5 $& $\pm 4.0 $\\
     $150-185$  & 9.9  & $+2.3, -1.9$ & $\pm 0.3 $& $\pm 2.8 $\\
     $185-250$  & 3.1  & $+1.3, -0.9$ & $\pm 0.1 $& $\pm 1.5 $\\
     $\geq$250  & 1.0  & $+0.8, -0.5$ & $\pm 0.1 $& $\pm 0.8 $\\

     \hline
     \end{tabular}
     \label{tab:QCDerrors}
\end{table}

Uncertainties in the EWK background prediction
include the statistical uncertainty from the $\Pe\Pgg$ control
sample and the 30\% uncertainty in the rate at which
electrons are misidentified as photons. The statistical
uncertainty is less than 9\% in each of the six signal bins.

There are also several uncertainties associated with
the signal efficiency. The statistical uncertainty
from the size of the T5gg or T6gg signal scans ranges
from 2 to 44\% depending on the mass bin. The
PDF uncertainties in the cross sections for signal simulation
are between 19 and 35\% and are taken from Ref.~\cite{Borschensky:2014cia}.
Other uncertainties include how well the jet energy
scale is known (1 to 30\%) and the uncertainty in
the photon identification efficiency (2.5\%).
The uncertainty in the integrated luminosity of the data sample is 2.5\%~\cite{CMS-PAS-LUM-17-001}.

\section{Results}
\label{sec:results}
We determine 95\% confidence
level (\CL) upper limits on gluino pair production and squark
pair production cross sections using the modified frequentist \CLs
method~\cite{Junk:1999kv, Read:2002hq}. The test statistic is an
LHC-style profile likelihood ratio~\cite{CMS-NOTE-2011-005}, and its
distribution is determined using
the asymptotic approximation~\cite{Cowan:2010js}.
The likelihood function is constructed from the background and signal \ptmiss distributions
across the six bins described in Section~\ref{sec:EstBkg}. The systematic uncertainties
described in Section~\ref{sec:SysUncert} are included in the test statistic as
constrained nuisance parameters. Systematic uncertainties which 
directly affect the yields of processes are assumed to follow a 
log-normal probability distribution, while statistical
uncertainties from the limited size of the control samples and
the signal MC samples are modeled using gamma probability distributions.

\begin{figure}[htb!]
\centering
 \includegraphics[width=0.6\textwidth]{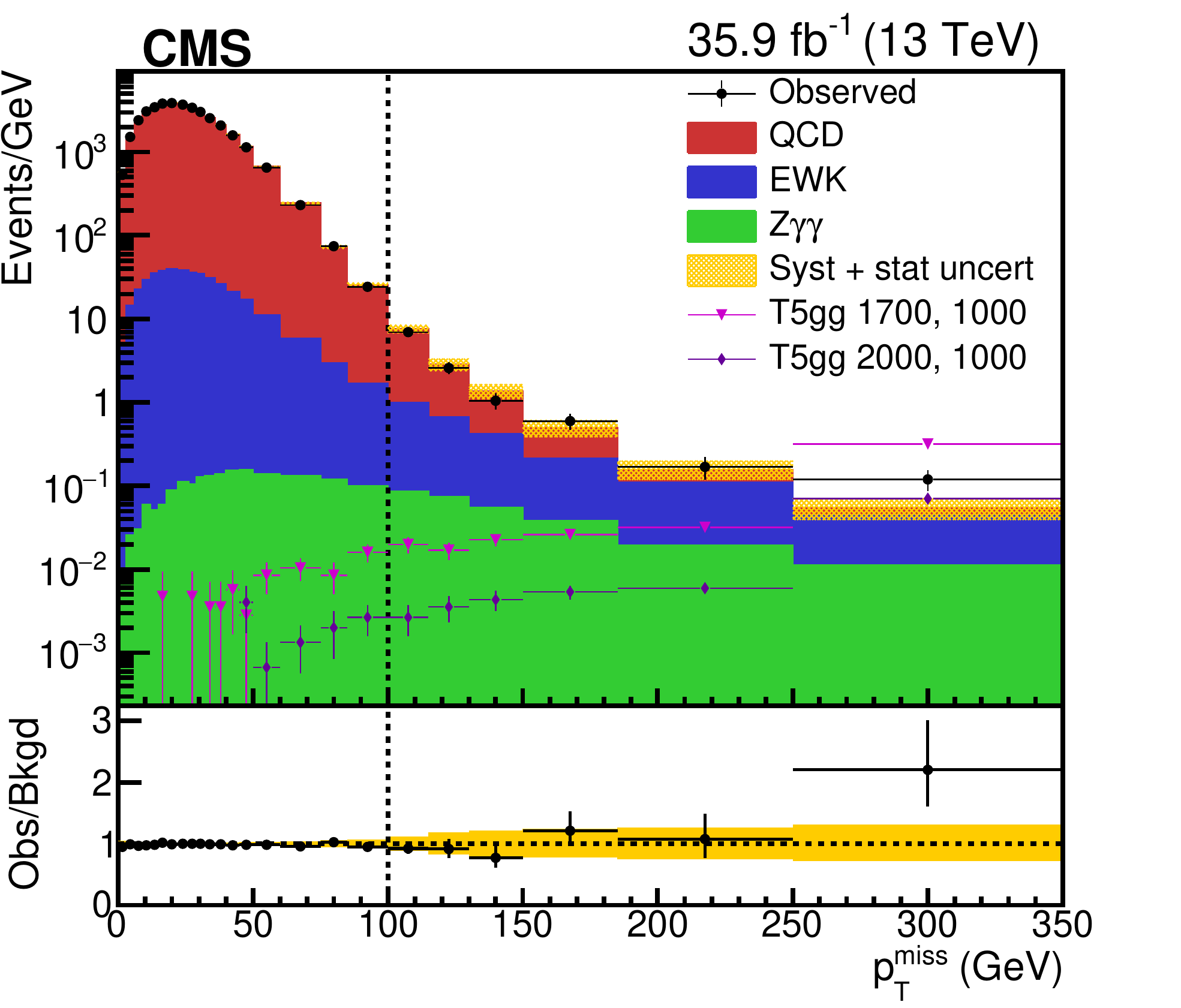}
 \caption{The top panel shows the observed
\ptmiss distribution in data (black points) and predicted
background distributions prior to the fit described in the text. The vertical line marks the boundary
between the control region ($\ptmiss < 100$\GeV) and the signal region ($\ptmiss > 100$\GeV).
The last bin is the overflow bin and includes all events with $\ptmiss > 250$\GeV.
The QCD background is shown in red, the EWK background
is shown in blue, and the $\PZ\Pgg\Pgg$ background is shown in green.
The \ptmiss distribution shown in pink (purple) corresponds to the T5gg
simplified model with $m_{\PSg}=1700$ (2000)\GeV and $m_{\PSGczDo} = 1000$\GeV.
The \ptmiss distributions from the T6gg simplified model are similar to the T5gg 
distributions shown here.
The bottom panel shows the ratio of observed events to the expected background.
The error bars on the ratio correspond to the
statistical uncertainty in the number of observed events.
The shaded region corresponds
to the total uncertainty in the background estimate. }
 \label{fig:Final}
\end{figure}

The full background prediction
and the measured \ptmiss distribution prior to the fit are shown in Fig.~\ref{fig:Final}.
The expected and observed numbers of events for each bin in the signal region 
are shown in Table~\ref{tab:ExpObs} for the pre-fit distributions and Table~\ref{tab:ExpObsv2}
for the post-fit distributions. 
Notably, in the last bin we observe 12 events and expect ${ 5.4 }^{+ 1.6 }_{- 1.5 }$
background events (pre-fit). The significance of the observed data after the fit across all
six bins of the signal region is calculated using the likelihood ratio test 
for each mass pair value of $m_{\PSGczDo}$ versus
$m_{\PSg}$ or $m_{\PSGczDo}$ versus $m_{\PSQ}$ for the T5gg and T6gg models, respectively.
The significance does not strongly depend on the SUSY masses, and 
for all masses in both models, the significance is found to correspond to
between 2.35 and 2.45 standard deviations. 
Several studies were performed to characterize the fit and the excess in the
final \ptmiss bin and to ensure that the statistical treatment of the data
is robust. In particular, the pre- and postfit distributions
were checked to make sure that the pulls are
consistent within the uncertainties.

\begin{table}[ht]
    \centering
    \topcaption{Number of expected background and observed data events 
    with 35.9\fbinv of 13\TeV data
    in the signal region prior to the fit defined in the text.
    The uncertainty in each expected background yield includes the 
    statistical uncertainty and all of the systematic uncertainties described in
        Section~\ref{sec:SysUncert} added in quadrature.}
    \begin{tabular}{ c c c c c c}
        \hline
        \ptmiss bin (\GeVns{}) & QCD & EWK &  $\PZ\Pgg\Pgg$ & Total background & Observed \\ [0.5ex]
        \hline
        $100-115$       & $99 \pm 12$                 & $13.7 \pm 4.2$  & $1.3 \pm 0.6$       & $ 114 \pm 13$                  & 105 \\
        $115-130$       & ${32.8}^{+ 7.0}_{- 6.7}$      & $ 9.0 \pm 2.7$  & $1.1 \pm 0.6$       & ${ 42.9 }^{+ 7.5 }_{- 7.3 }$    & 39 \\
        $130-150$       & ${18.8}^{+ 5.1}_{- 4.9}$      & $ 7.4 \pm 2.3$  & $1.1 \pm 0.6$       & ${ 27.3 }^{+ 5.6 }_{- 5.4 }$    & 21 \\
        $150-185$       & ${9.9}^{+ 3.6}_{- 3.4}$       & $ 6.1 \pm 1.9$  & $1.3 \pm 0.7$       & ${ 17.4 }^{+ 4.1 }_{- 3.9 }$    & 21 \\
        $185-250$       & ${3.1}^{+ 1.9}_{- 1.7}$       & $ 5.8 \pm 1.8$  & $1.3 \pm 0.6$       & ${ 10.2 }^{+ 2.7 }_{- 2.6 }$    & 11 \\
        $\geq$250       & ${1.0}^{+ 1.1}_{- 0.9}$       & $ 3.3 \pm 1.1$  & $1.1 \pm 0.6$       & ${ 5.4 }^{+ 1.6 }_{- 1.5 }$     & 12\\

        \hline
    \end{tabular}
    \label{tab:ExpObs}
\end{table}

\begin{table}[ht]
    \centering
    \topcaption{Number of expected background and observed data events 
    with 35.9\fbinv of 13\TeV data in the signal region after the fit defined in the text.
        The stated uncertainties are the post-fit uncertainties in each expected background yield.}
    \begin{tabular}{ c c c c c c}
        \hline
        \ptmiss bin (\GeVns{}) & QCD & EWK &  $\PZ\Pgg\Pgg$ & Total background & Observed \\ [0.5ex]
        \hline
        $100-115$       & $92.7 \pm 7.9$                & $15.9 \pm 3.8$ & $1.6 \pm 0.8$       & $ 110.1 \pm 7.4$                 & 105 \\
        $115-130$       & $29.7 \pm 4.4$                & $10.4 \pm 2.5$ & $1.4 \pm 0.7$       & $ 41.5 \pm 3.9$                  &  39 \\
        $130-150$       & $16.0 \pm 3.2$                & $8.5 \pm 2.1$  & $1.3 \pm 0.7$       & $ 25.9 \pm 3.1$                  &  21 \\
        $150-185$       & $9.3 \pm 2.7$                 & $7.1 \pm 1.8$  & $1.6 \pm 0.8$       & $ 18.1 \pm 2.6$                  &  21 \\
        $185-250$       & $2.6 \pm 1.2$                 & $6.7 \pm 1.6$  & $1.6 \pm 0.8$       & $ 10.9 \pm 1.8$                  &  11 \\
        $\geq$250       & $0.7 \pm 0.8$                 & $4.0 \pm 1.0$  & $1.4 \pm 0.7$       & $ 6.0 \pm 1.2$                   &  12 \\
        \hline
    \end{tabular}
    \label{tab:ExpObsv2}
\end{table}

\begin{figure}[htbp]
    \centering
    \includegraphics[width=0.49\textwidth]{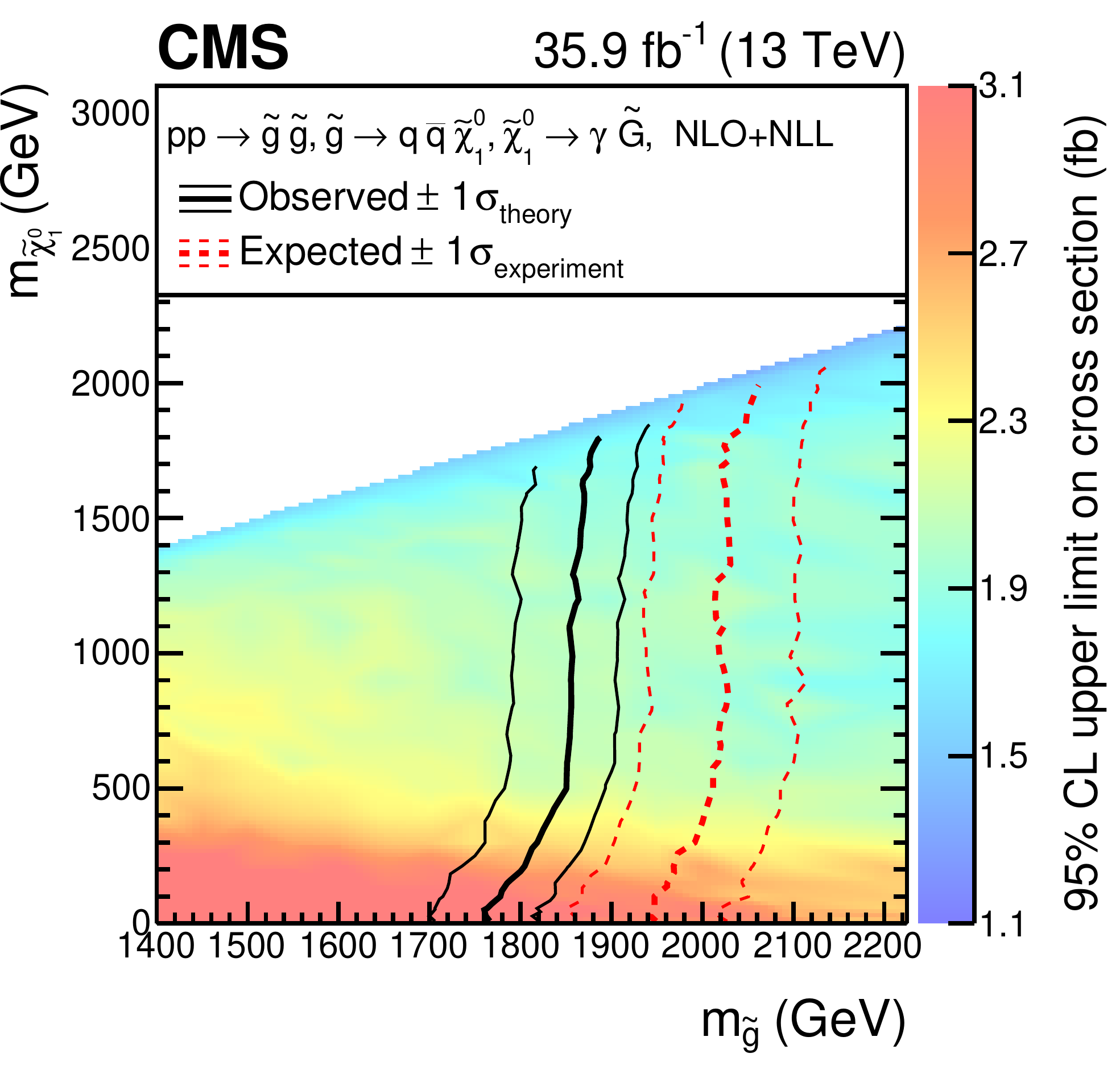}
    \includegraphics[width=0.49\textwidth]{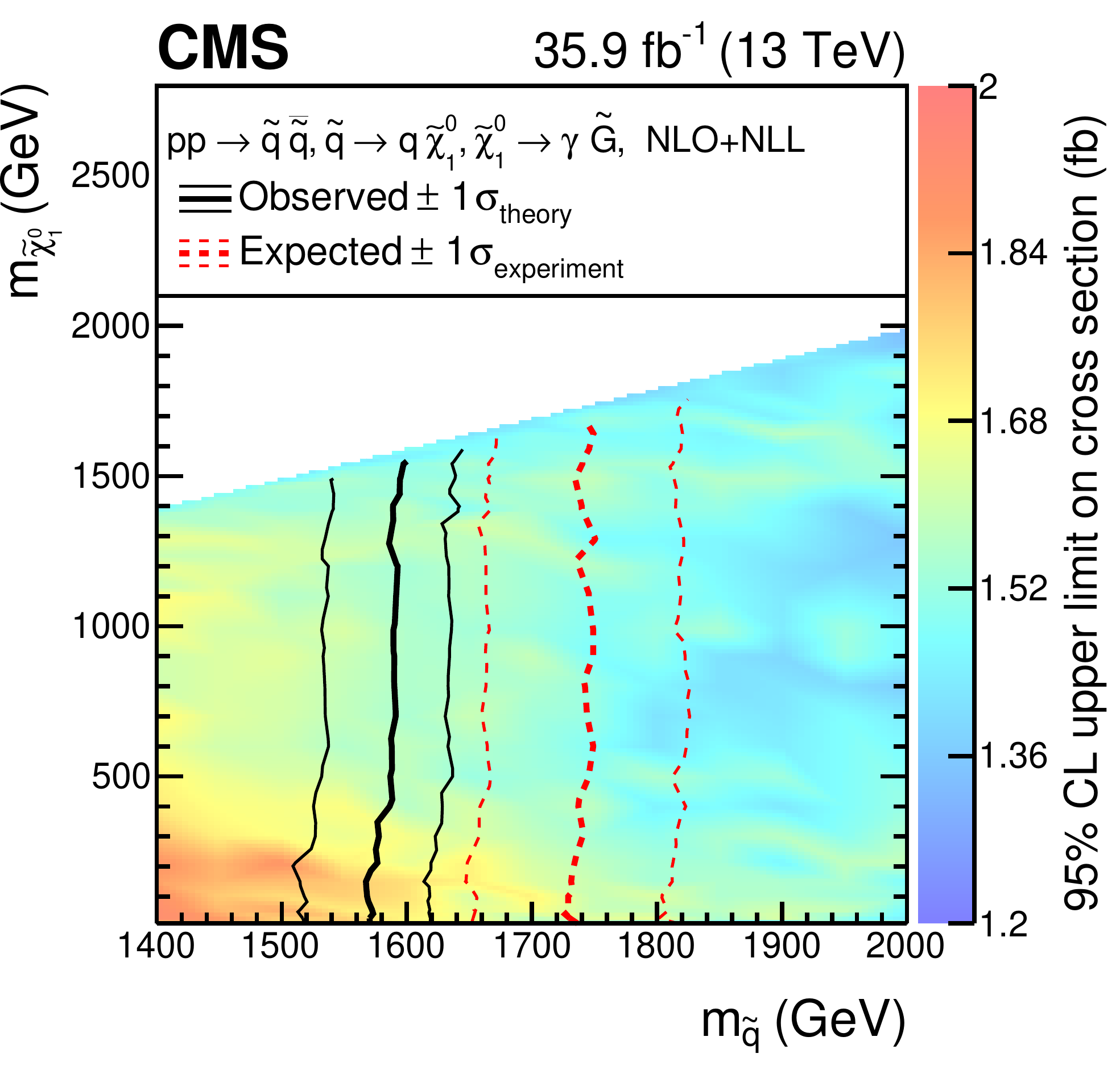}
    \caption{The 95\% confidence level upper limits on the gluino (left) and squark (right) pair
        production cross sections as a function of gluino or squark and neutralino masses.
        The contours show the observed and expected exclusions assuming
        the NLO+NLL cross sections, with their one standard deviation
        uncertainties.}
    \label{fig:limit}
\end{figure}

In Fig.~\ref{fig:limit} we present 95\% \CL upper limits on the gluino and
squark pair production cross sections as a function of the mass pair values 
for the two models considered in this analysis.
From the NLO+NLL
predicted signal cross sections and their uncertainties we derive contours
representing lower limits in the SUSY mass plane. We also show expected limit
contours based on the expected experimental cross section limits and their
uncertainties. For values of the neutralino mass between 500 and 1500\GeV,
we expect to exclude
gluino masses up to 2.02\TeV and squark masses up to 1.74\TeV. This
is an improvement of approximately 400 and 300\GeV, respectively,
upon the reach of the previous CMS result~\cite{CMS:2016_anal}.
We observe exclusions for gluino masses up to 1.86\TeV and squark masses
up to 1.59\TeV. The observed exclusions are
lower than the expected exclusions because of the observed excess in the data.

\section{Summary}
\label{sec:summary}
The results of a search for general
gauge-mediated supersymmetry breaking
in proton-proton collisions with two photons
and missing transverse momentum
in the final state are reported.
The analysis was performed using data corresponding to 35.9\fbinv
of integrated luminosity, recorded with the CMS detector in
2016 at a proton-proton center-of-mass energy of 13\TeV.
An excess of events corresponding to 2.4 standard deviations is observed.
Limits are determined on the
masses of supersymmetric particles in two
simplified models using
data-driven background estimation methods and
NLO+NLL signal cross section calculations.

In both models, the next-to-lightest supersymmetric
particle is the neutralino, which decays with a 100\%
branching fraction to a photon and a gravitino,
the lightest supersymmetric particle. The first
simplified model assumes gluino pair production,
with each gluino decaying to a neutralino and quarks.
The second simplified model
assumes squark pair production, with each squark
decaying to a quark and a neutralino.
The expected limits on gluino and squark masses, for the respective
models, are 2.02 and 1.74\TeV at 95\% confidence level. This is an
increase in sensitivity of more than 300\GeV for each model with
respect to the analysis performed with 2.3\fbinv of integrated luminosity
collected using the CMS detector in 2015.
The observed exclusions are for gluino
masses less than 1.86\TeV and squark masses
less than 1.59\TeV, where the difference
between the expected and observed exclusions is
driven by the excess observed in the data. The analysis described
in this paper improves the observed limits by
210\GeV for gluino masses and
220\GeV for squark masses with
respect to the previous CMS result.

\begin{acknowledgments}
We congratulate our colleagues in the CERN accelerator departments for the excellent performance of the LHC and thank the technical and administrative staffs at CERN and at other CMS institutes for their contributions to the success of the CMS effort. In addition, we gratefully acknowledge the computing centers and personnel of the Worldwide LHC Computing Grid for delivering so effectively the computing infrastructure essential to our analyses. Finally, we acknowledge the enduring support for the construction and operation of the LHC and the CMS detector provided by the following funding agencies: BMBWF and FWF (Austria); FNRS and FWO (Belgium); CNPq, CAPES, FAPERJ, FAPERGS, and FAPESP (Brazil); MES (Bulgaria); CERN; CAS, MoST, and NSFC (China); COLCIENCIAS (Colombia); MSES and CSF (Croatia); RPF (Cyprus); SENESCYT (Ecuador); MoER, ERC IUT, and ERDF (Estonia); Academy of Finland, MEC, and HIP (Finland); CEA and CNRS/IN2P3 (France); BMBF, DFG, and HGF (Germany); GSRT (Greece); NKFIA (Hungary); DAE and DST (India); IPM (Iran); SFI (Ireland); INFN (Italy); MSIP and NRF (Republic of Korea); MES (Latvia); LAS (Lithuania); MOE and UM (Malaysia); BUAP, CINVESTAV, CONACYT, LNS, SEP, and UASLP-FAI (Mexico); MOS (Montenegro); MBIE (New Zealand); PAEC (Pakistan); MSHE and NSC (Poland); FCT (Portugal); JINR (Dubna); MON, RosAtom, RAS, RFBR, and NRC KI (Russia); MESTD (Serbia); SEIDI, CPAN, PCTI, and FEDER (Spain); MOSTR (Sri Lanka); Swiss Funding Agencies (Switzerland); MST (Taipei); ThEPCenter, IPST, STAR, and NSTDA (Thailand); TUBITAK and TAEK (Turkey); NASU and SFFR (Ukraine); STFC (United Kingdom); DOE and NSF (USA).

\hyphenation{Rachada-pisek} Individuals have received support from the Marie-Curie program and the European Research Council and Horizon 2020 Grant, contract Nos.\ 675440 and 765710 (European Union); the Leventis Foundation; the A.P.\ Sloan Foundation; the Alexander von Humboldt Foundation; the Belgian Federal Science Policy Office; the Fonds pour la Formation \`a la Recherche dans l'Industrie et dans l'Agriculture (FRIA-Belgium); the Agentschap voor Innovatie door Wetenschap en Technologie (IWT-Belgium); the F.R.S.-FNRS and FWO (Belgium) under the ``Excellence of Science -- EOS'' -- be.h project n.\ 30820817; the Beijing Municipal Science \& Technology Commission, No. Z181100004218003; the Ministry of Education, Youth and Sports (MEYS) of the Czech Republic; the Lend\"ulet (``Momentum'') Programme and the J\'anos Bolyai Research Scholarship of the Hungarian Academy of Sciences, the New National Excellence Program \'UNKP, the NKFIA research grants 123842, 123959, 124845, 124850, 125105, 128713, 128786, and 129058 (Hungary); the Council of Science and Industrial Research, India; the HOMING PLUS program of the Foundation for Polish Science, cofinanced from European Union, Regional Development Fund, the Mobility Plus program of the Ministry of Science and Higher Education, the National Science Center (Poland), contracts Harmonia 2014/14/M/ST2/00428, Opus 2014/13/B/ST2/02543, 2014/15/B/ST2/03998, and 2015/19/B/ST2/02861, Sonata-bis 2012/07/E/ST2/01406; the National Priorities Research Program by Qatar National Research Fund; the Programa Estatal de Fomento de la Investigaci{\'o}n Cient{\'i}fica y T{\'e}cnica de Excelencia Mar\'{\i}a de Maeztu, grant MDM-2015-0509 and the Programa Severo Ochoa del Principado de Asturias; the Thalis and Aristeia programs cofinanced by EU-ESF and the Greek NSRF; the Rachadapisek Sompot Fund for Postdoctoral Fellowship, Chulalongkorn University and the Chulalongkorn Academic into Its 2nd Century Project Advancement Project (Thailand); the Welch Foundation, contract C-1845; and the Weston Havens Foundation (USA).
\end{acknowledgments}
\bibliography{auto_generated}

\cleardoublepage \appendix\section{The CMS Collaboration \label{app:collab}}\begin{sloppypar}\hyphenpenalty=5000\widowpenalty=500\clubpenalty=5000\vskip\cmsinstskip
\textbf{Yerevan Physics Institute, Yerevan, Armenia}\\*[0pt]
A.M.~Sirunyan, A.~Tumasyan
\vskip\cmsinstskip
\textbf{Institut f\"{u}r Hochenergiephysik, Wien, Austria}\\*[0pt]
W.~Adam, F.~Ambrogi, E.~Asilar, T.~Bergauer, J.~Brandstetter, M.~Dragicevic, J.~Er\"{o}, A.~Escalante~Del~Valle, M.~Flechl, R.~Fr\"{u}hwirth\cmsAuthorMark{1}, V.M.~Ghete, J.~Hrubec, M.~Jeitler\cmsAuthorMark{1}, N.~Krammer, I.~Kr\"{a}tschmer, D.~Liko, T.~Madlener, I.~Mikulec, N.~Rad, H.~Rohringer, J.~Schieck\cmsAuthorMark{1}, R.~Sch\"{o}fbeck, M.~Spanring, D.~Spitzbart, W.~Waltenberger, J.~Wittmann, C.-E.~Wulz\cmsAuthorMark{1}, M.~Zarucki
% ...
\vskip\cmsinstskip
% ...
1:  Also at Vienna University of Technology, Vienna, Austria\\
% ...
\end{sloppypar}
\end{document}